\documentclass[10pt]{article}
\usepackage[OE]{express}
\begin{document}

\title{Two-Photon Lensless Micro-endoscopy with \textit{in-situ} Wavefront Correction}

\author{Uri Weiss,\authormark{1} and Ori Katz,\authormark{1,*} }

\address{\authormark{1}Department of Applied Physics, The The Hebrew University of Jerusalem\\}

\email{\authormark{*}orik@mail.huji.ac.il} 

\begin{abstract*}
\textbf{Multi-core fiber-bundle endoscopes provide a minimally-invasive solution for deep tissue imaging and opto-genetic stimulation, at depths beyond the reach of conventional microscopes. 
Recently, wavefront-shaping has enabled lensless bundle-based micro-endoscopy by correcting the wavefront distortions induced by core-to-core inhomogeneities. However, current wavefront-shaping solutions require access to the fiber distal end for determining the bend-sensitive wavefront-correction. 
Here, we show that it is possible to determine the wavefront correction \textit{in-situ}, without any distal access. Exploiting the nonlinearity of two-photon excited fluorescence, we adaptively determine the wavefront correction \textit{in-situ}, using only proximal detection of epi-detected fluorescence. We experimentally demonstrate diffraction-limited, three-dimensional, two-photon lensless microendoscopy with commercially-available ordered- and disordered multi-core fiber bundles. }
\end{abstract*}

\section{Introduction}
Flexible optical endoscopes are an important tool for a variety of applications from clinical procedures to biomedical investigations. A common use of such endoscopes is macroscopic imaging inside hollow organs. Another important application is in micro-endoscopy, where micron-scale structures such as single neurons are imaged or optically excited at depths beyond the reach of conventional microscopes. The latter are limited in their penetration depth due to tissue scattering and absorption \cite{Vasilis2010GoingDeeper}. 

In recent years, various solutions for small diameter micro-endoscopes have been developed \cite{flusberg2005fiberReviewNature,oh2013fiberReview}: Micro-endoscopes that are based on single-mode fibers are bend-insensitive, but require distal optical elements such as scanners and lenses \cite{flusberg2005fiberReviewNature}, or spectral dispersers \cite{Sylwia2015Scanning-free,Barankov2014HighThrough-put} to produce an image. Such distal elements may significantly enlarge the endoscope's diameter, increasing tissue damage, and consequently limiting its use for deep-tissue imaging. Developing a flexible, lensless micro-endoscope with a minimal diameter is thus a sought after goal for minimally-invasive deep-tissue imaging.

Currently, the solutions for constructing lensless endoscopes are based either on imaging fiber bundles \cite{flusberg2005fiberReviewNature} or wavefront-shaping \cite{warren2016adaptiveEndosMicroscope, andresen2013twoPhotoneLenslessEndoscope,Thompson2011Adaptive,bianchi2012multi,vcivzmar2012exploiting}. Imaging fiber bundles consist of thousands of single-mode cores packed together, where each core functions as a single pixel. While common and straightforward to use, conventional lensless bundles-based endoscopes suffer from limited resolution, pixelation, poor axial sectioning, and a small and fixed working distance. 

While axial-sectioning can be obtained in fiber bundles by addition of confocal scanning \cite{flusberg2005fiberReviewNature} or structured illumination\cite{Nenad2008FluorescenceEndomicroscopy}, the working distance is fixed to the fiber facet or its image, and does not allow three-dimensional (3D) imaging.

In recent years, a number of works demonstrated the use of wavefront-shaping for microendoscopy \cite{mosk2012controlling,bianchi2012multi,vcivzmar2012exploiting,papadopoulos2012focusing,choi2012scanner,andresen2013twoPhotoneLenslessEndoscope,warren2016adaptiveEndosMicroscope,ploschner2015seeingThroughChaos}. 
In wavefront shaping, a computer-controlled spatial light modulator (SLM) is used to compensate the phase randomization and mode-mixing in fiber bundles or multi-mode fibers, in a principle similar to the decades-old works in holography \cite{Spitz1967Tranmission,Yariv1976Direct}. 
However, since the phase distortions in long fibers are sensitive to fiber bending, even with the state-of-the-art digital wavefront control a direct feedback from the fiber distal end or precise knowledge of the fiber shape \cite{ploschner2015seeingThroughChaos} are still required to determine the wavefront correction. 
These requirements make the application of wavefront-shaping techniques for \textit{flexible} endoscopes very challenging in most practical imaging scenarios.
 Recently, speckle correlations in the optical transmission-matrices of fiber bundles, known as 'memory effect' correlations, have been exploited for computationally reconstructing distal images from proximal measurements, without wavefront correction \cite{porat2016widefield,Nicolino2016CAlibration-free}. However, these approaches currently allow only two-dimensional imaging, and provide high-fidelity reconstruction only for simple objects. 

Here, we present a lensless two-photon microendoscope based on wavefront-shaping, which does not require any a-priori knowledge of the fiber transmission-matrix or access to the distal end, even after fiber bending. In our approach the wavefront correction is performed \textit{in-situ}, by iterative optimization of two-photon fluorescence (2PF) signals detected at the proximal end. 
Our approach is based on the fact that iterative optimization of a nonlinear signal leads to diffraction limited focusing \cite{katz2014noninvasive}, even when the signal is collected by a spatially-integrating detector having no spatial resolution.
This principle has been recently exploited for focusing light through scattering samples \cite{katz2014noninvasive} or multi-mode fibers\cite{rosen2015focusing2PFThroughMMF}, and here we show its use for endoscopic imaging.
Our imaging technique is composed of two steps: the first is focusing of the excitation beam on the target object by iterative wavefront optimization. After the wavefront-correction has been found, two-photon imaging is performed by scanning the formed focus in three-dimensions using the single wavefront correction, exploiting the 'memory-effect' of imaging fiber bundles \cite{porat2016widefield,  Nicolino2016CAlibration-free, Thompson2011Adaptive}.

\section{Methods}
\label{sec:Methods}
\begin{figure}[tbp]
\centering
\includegraphics[width=10cm]{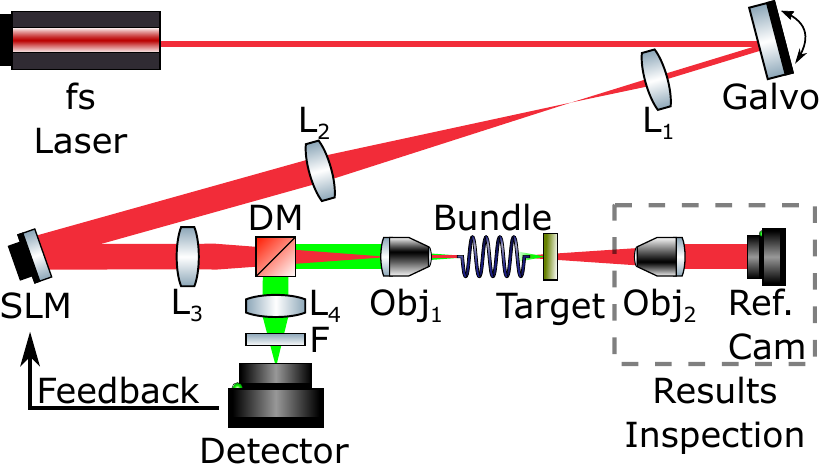}
\caption{Setup: 85 fs long laser pulses are reflected off a galvanomteric mirror and imaged on an SLM. The SLM is imaged on the proximal facet of a fiber-bundle. Two-photon fluorescent (2PF) targets are placed at short distances from the distal end of the fiber. The excited 2PF is collected by the same fiber and detected at the proximal end using a PMT or an sCMOS camera. The detected 2PF signal is used as a feedback for an iterative wavefront-shaping optimization process, aimed at maximizing the total detected 2PF, forming a sharp focus on the target. 3D two-photon imaging is achieved by scanning the formed focus with the SLM and galvanometric mirror. A reference camera is used only for results inspection.} 
\label{fig:setup}
\end{figure}

The experimental setup is illustrated in Fig.\ref{fig:setup}. A beam from a Ti:Sapphire laser oscillator (Mai-Tai DeepSee eHP, Spectra-Physics) producing 85 fs pulses at 785nm wavelength, is reflected off a galvanometric mirror (GVS012, Thorlabs), The galvanometric mirror is imaged on a phase-only SLM (X13138-02, Hamamatsu Photonics) by a 4-f telescope ($L_1 = 35 mm , L_2 = 500 mm$). The SLM is imaged on the proximal facet of a fiber bundle (either Schott 1533385 in Figs. 2-4, or Fujikura FIGH-03-215S in Figure \ref{fig:Fujikura}d,i) using an additional telescope ($L_3 = 250 [mm] , Obj_1;$ 20X Plan Achromat, Olympus). Two-photon fluorescent (2PF) target objects were placed simultaneously at several distances of 0.8mm-3mm from the fiber distal end. The excited 2PF signals are collected by the same bundle, propagate back to the proximal end, separated from the laser wavelength by a dichroic mirror (FF605-Di02-25x36, Semrock), and focused on a detector (either an sCMOS camera (Zyla 4.2 Plus, Andor) or a photomultiplier tube, PMT (X13138-02, Hamamtasu)), after additional filtering (FF01-510/84-25 bandpass, FF01-650/SP-25 short-pass filter, Semrock).  
2PF imaging is performed \textit{in-situ} in two steps: first, the SLM wavefront-correction for focusing is found by iteratively optimizing the total 2PF signal collected by the detector. This step produces a sharp focused beam on the target object \cite{katz2014noninvasive}. After the wavefront correction is found, the focused spot is raster-scanned in 3D by adding a parabolic phase on the SLM for axial scanning, and a wavefront tilt with the SLM and/or Galvo for lateral scanning. The 2PF signal detected at the proximal end is used to generate a 3D image of the target objects, as in conventional 2PF microscopy \cite{katz2014noninvasive}.
For inspecting the focusing performance, a reference camera (Mako U-130B, Allied Vision) is used to directly image the object plane, using a microscope objective ($Obj_2;$ 10X Plan Achromat, Olympus). Importantly, the reference camera is used only for results inspection, and is not required for successful focusing or imaging.

\section{Results}
\label{sec:Results}
\subsection{In-situ Focusing}

\begin{figure}[tbp]
\centering
\includegraphics[width=11cm]{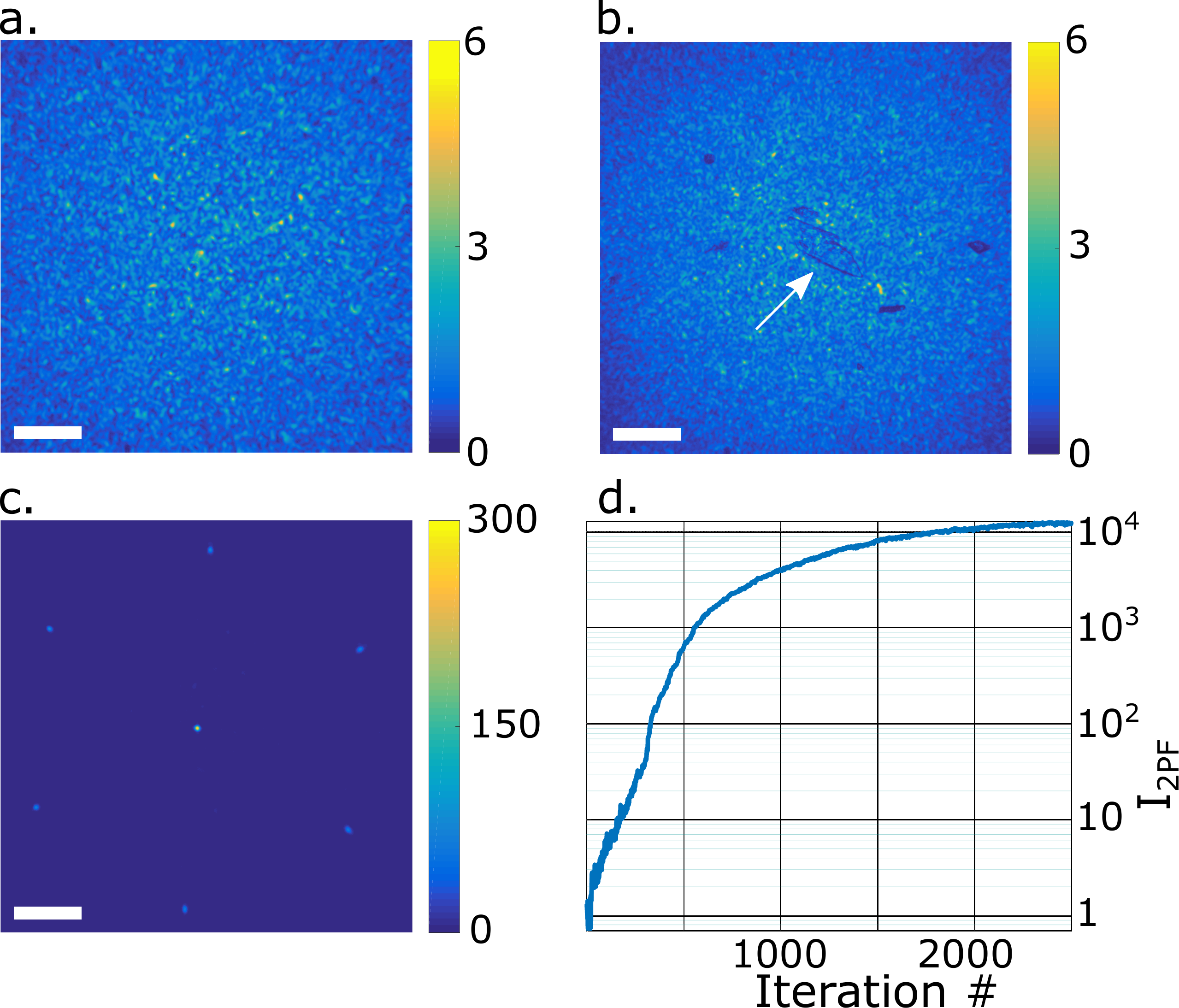}
\caption{Experimental focusing through a fiber-bundle with proximal-only detection. (a) The speckle pattern at the object plane (z=2.2mm) before optimization. (b) Same as (a), with a 2PF object in the field of view (marked by arrow).  (c) Same as (a) after optimization of the epi-detected 2PF, showing sharp focusing. The side-lobes are a result of the lattice periodicity of the bundle cores. (d) Evolution of the proximally-detected 2PF during the optimization process. Scale Bars: 100 $\mu m$. (color-bars are normalized such that the mean intensity in (a) is 1).}
\label{fig:optimization}
\end{figure}

As a first demonstration, 2PF particles of Coumarin 307 were placed at distances of $ 2mm-3mm$ from the bundle's distal facet, at two axial planes. 
Initially, the light at the object plane, as recorded by the reference camera, was a random speckle pattern (Fig.\ref{fig:optimization}a). After 2,500 iterations of an optimization algorithm maximizing the total collected 2PF signal, $I_{2PF}$, a sharp focal spot was formed on the object closest to the fiber facet (Fig.\ref{fig:optimization}c). The obtained  focus size is of the speckle-grain (diffraction-limited) dimensions (see Fig.4), and has a peak to background ratio (PBR) of $\sim 310$  (Fig.\ref{fig:optimization}c).  For optimizing the 2PF signal we have used an iterative partitioning algorithm \cite{Vellekop2008PhaseControlAlgorithms,Katz2011FocusingandCompression}. In this algorithm, the SLM was divided into 128X160 equally sized square segments. In each iteration, a phase, $\phi_n$  of zero to $2\pi$ is added to a random subset of these segments in $N=4$ steps. The 2PF signal as a function of the added phase is fitted to a cosine: $I_{2PF}(\phi_n)=A+B\cdot cos(\phi_n - \phi)$, using a fast Fourier transform, and the phase $\phi$ that  maximizes the 2PF signal is added to the chosen segments.  The obtained 2PF signal as a function of the iteration is plotted in Figure \ref{fig:optimization}d.  The total optimization time was limited by the refresh rate of the liquid-crystal SLM (~5Hz) to tens of minutes. However, the optimization time could be significantly reduced since the integration times were of the order of   $0.1[ms]$  even in the first iterations of the optimization (see Discussion). 

\subsection{3D two-photon imaging}
\begin{figure}[tbp]
\centering
\includegraphics[width=10cm]{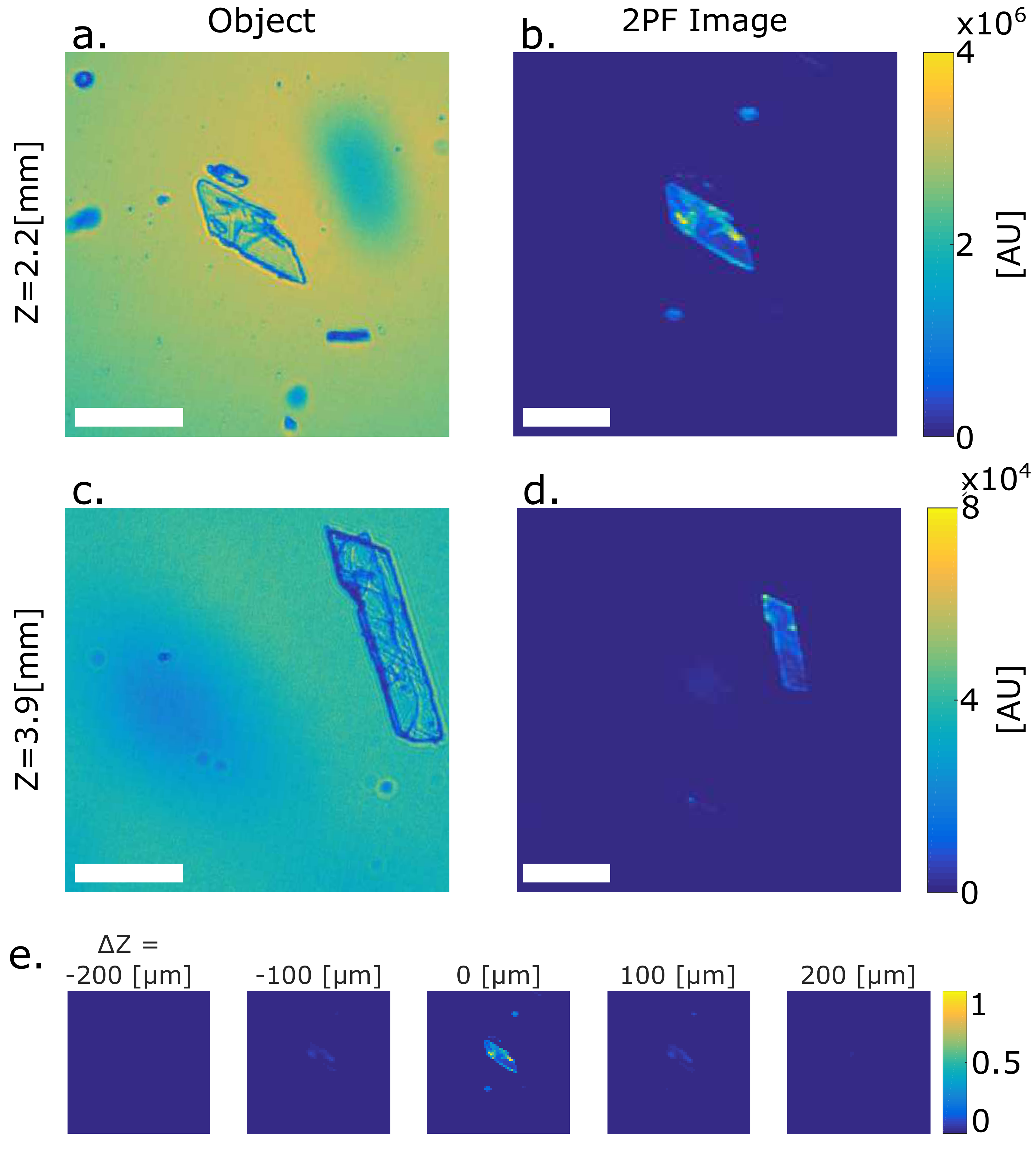}
\caption{Lensless two-photon imaging obtained by scanning the focal spot generated \textit{in-situ} (Fig.1c). (a,c) Reference camera bright-field images of fluorescent objects placed simultaneously at two axial planes. (b,d) Corresponding two-photon images obtained through the bundle. (e) Two-photon images at various focal planes near the plane of (b), demonstrating axial sectioning.
Scale Bars: (a,c) 100 $\mu m$; (b,d) 0.05 radians, (the farthest object has a smaller \textit{angular} extent).}
\label{fig:Reconstruction}
\end{figure}

Following the adaptive focusing, 3D imaging was performed by raster scanning the focus. Fast scanning, with a pixel dwell-time significantly shorter than the SLM refresh rate, was achieved by using a galvanometric mirror for fast scanning in one lateral dimension. Scanning in the other lateral dimension was implemented by addition of a linear phase ramps  on the SLM wavefront correction. Scanning in the axial dimension was achieved by adding parabolic phase patterns to the SLM wavefront correction, effectively transforming the fiber bundle to an adaptive lens. Fig.\ref{fig:Reconstruction}(a-d) shows the 2PF images obtained with the focus scanning for two axial planes, where the fluorescent objects were present. Thanks to the inherent axial sectioning of 2PF focused excitation, at each plane, only the objects residing in this plane are visible (Fig.\ref{fig:Reconstruction}b,d), whereas the parts of the object residing at different axial planes do not contribute to a substantial background halo, as is visible in the conventional bright-field imaging performed with the reference camera (Fig.\ref{fig:Reconstruction}a,c.). Fig.\ref{fig:Reconstruction}(e) shows additional two-photon images obtained at axial planes close to the plane of Fig.\ref{fig:Reconstruction}(a-b).

\subsection{Characterization of the formed focus}
\begin{figure}[t!p]
\centering
\includegraphics[width=10cm]{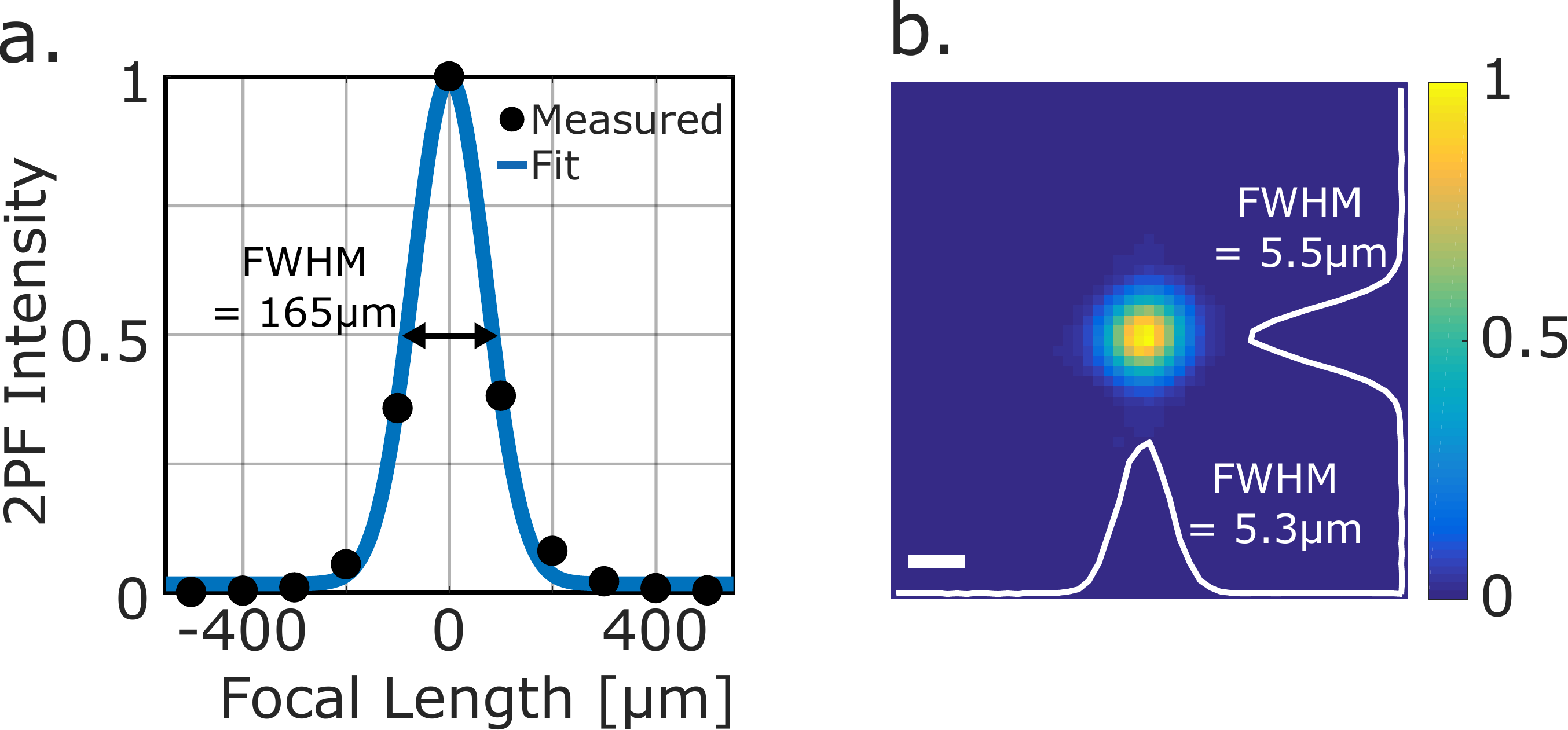}
\caption{Focal spot characterization: (a) Axial characterization of the focus width obtained by scanning the focal spot in the z-direction on a thin 2PF object, by adding a parabolic phase to the wavefront-correction. (b) Lateral characterization using the reference camera. Measured FWHMs are $5.3\pm0.1$ and $5.5\pm0.1 \mu m $ in the x- and y-axis respectively.  Scale Bar, 5 $\mu m$.}
\label{fig:Profile}
\end{figure}

In wavefront shaping, the focusing resolution is expected to be diffraction-limited, as dictated by the dimensions of a single speckle grain \cite{vellekoop2007focusing}
To characterize the axial resolution, the 2PF signal in the image stack presented in Fig.\ref{fig:Reconstruction}(e) was plotted as a function of the axial distance. Fig.\ref{fig:Profile}a displays the trace of the maximal 2PF intensity obtained at each depth. The axial resolution, defined by the FWHM of a fit to a Gaussian, is  $\delta z = 165\mu m\pm 10$. The axial resolution, $\delta z$ in 2PF microscopy is expected to be $0.64$  times the FWHM of the focused beam, which is twice its Rayleigh range: $FWHM_{2PF} = 0.64 \cdot 2\cdot Z_{R}=\frac{0.64 \cdot2\pi w_{0}^{2}}{\lambda/n}={\frac{0.64 }{ln\left(2\right)}}\frac{\pi\delta x ^2}{\lambda/n}$, where $w_0$ is the focused beam waist,  $\delta x$ is the resulting focus transverse FWHM, $\lambda=785 nm$  is the laser wavelength, and $n$ is the refractive index of the medium ($\sim1.45$ in this experiment).

To experimentally characterize the transverse dimensions of the focus, $\delta x$ , the fluorescent object was removed, and the focus was imaged by the reference camera.  The measured transverse (x-y) focal spot size was measured to have a FWHM of$~5.4 \mu m \pm 0.2$  (Fig.\ref{fig:Profile}(b)). A value which is in accordance with the measured axial resolution, and also the expected diffraction-limited spot size dictated by the measurement geometry \cite{porat2016widefield}: 
\begin{equation}
\delta x\approx\left[\left(\frac{\lambda}{D_{bundle}}\cdot Z\right)^{2}+\left(\frac{\lambda}{NA}\right)^{2}\right]^{\frac{1}{2}}	
\label{eq:SpecklSize}
\end{equation}
where $D_{bundle}\approx0.45mm$ is the fiber bundle's diameter, $Z\approx 2.2mm$ is the distance between the fiber facet and the object in this experiment, and $NA\approx 0.35$ is the numerical aperture.

\subsection{Comparison of ordered vs. disordered bundles for wavefront-shaping based endoscopy}

\begin{figure}[t!]
\centering
\includegraphics[width=10cm]{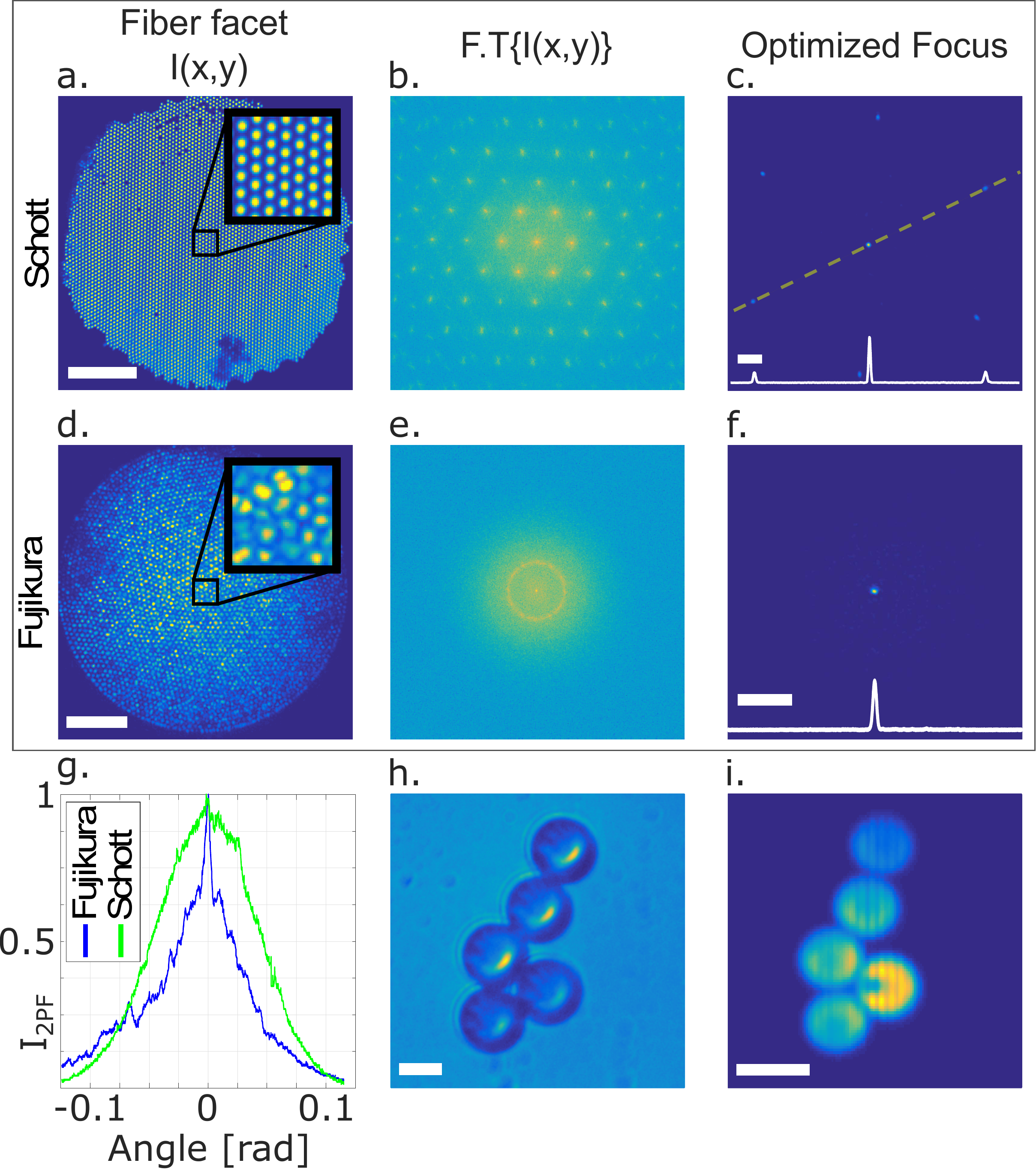}
\caption{Focusing with suppressed side-lobes using bundles with disordered cores. (a,b,c) Focusing results using a fiber bundle with ordered cores (Schott 1563385), used in Figs 1-4: (a) Image of the intensity distribution on the fiber distal facet, showing the cores ordered arrangement. (b) Fourier transform of (a). (c) optimized focal spot (white line: intensity cross-section along the dashed line). (d,e,f) Same as (a-c), using a bundle with disordered cores (Fujikura FIGH-03-215S), showing suppressed side-lobes. (g) Comparison of the angular scanning range (the 'memory effect') of the two fibers. The Fujikura fiber shows a smaller memory-effect range, likely due to light propagation between the cores (see (d)), resulting in a smaller FoV. (h,i) Two-photon imaging of fluorescent beads through the disordered fiber: (h) bright-field image of the object, taken with the reference camera. (i) two-photon image through the bundle obtained with the proposed approach. Scale Bars: (a,c,d,f) 50 $\mu m$, (h) 10 $\mu m$, (i) 0.02 rad.
}
\label{fig:Fujikura}
\end{figure}

As is visible in the focusing results of Fig.\ref{fig:optimization}(c), the obtained focal spot is surrounded by six visible side-lobes, positioned in a hexagonal pattern. These side-lobes are the result of the hexagonal periodicity of the cores in the fiber bundle. Fig.\ref{fig:Fujikura}a shows an image of the fiber facet  used in the above experiments, displaying the hexagonal lattice order. The side-lobes in the formed focus are predicted by the Fourier transform of the cores arrangement (Fig.\ref{fig:Fujikura}b,c) \cite{sivankutty2016AperiodicFibre,kim2018semi}. To suppress the side-lobes, we tested a commercially-available fiber with a less ordered arrangement of cores (Fujikura FIGH-03-215S). Indeed, using such a disordered fiber the side-lobes are effectively suppressed (Fig.\ref{fig:Fujikura}d-f). 

The main advantage of using non-ordered fiber bundles is the absence of diffraction side-lobes. When used for focus scanning based imaging, such side-lobes produce replicas of the object parts that are scanned by them. Thus, the side-lobes limit the transverse extent, i.e. field of view (FoV), of the imaged objects to be smaller than $x \approx z\cdot \lambda / d$  , where $d$ is the core-to-core spacing \cite{sivankutty2016AperiodicFibre}. Fig\ref{fig:Fujikura}h,i displays a 2PF image obtained with our approach using the disordered fiber, showing accurate diffraction-limited 2PF image of the target objects (fluorescent beads). 

While we have expected the disordered fiber to provide a considerably larger FoV, we have noticed that this was not the case, as is quantified by measuring the reduction in focus intensity as a function of the scan angle (Fig.5g). This narrower 'memory-effect' angular range of this specific fiber is not an inherent limitation of disordered fibers. We attribute the smaller FoV to light propagation \textit{between} the cores of this fiber. This can be observed in the lower cores-to-background contrast in Fig.5d, compared to Fig.5a. Another effect that this imperfect light guidance is causing is a narrower spectral speckle correlation bandwidth that we have measured for this fiber (not shown). 
This narrower spectral correlation bandwidth leads to a lower speckle contrast \cite{Curry2011DirectDetermination} that is measured when the 12nm-wide femtosecond pulsed illumination is used. This in turn, lowers the focus intensity enhancement obtained by wavefront shaping \cite{vellekoop2007focusing}. To increase the initial speckle contrast and focusing PBR using this fiber, a narrow BPF (LL01-785-25, Semrock, 3nm FWHM) was used in the illumination path in the experiments involving this fiber.

\section{Discussion}
\label{Discussion}

We have demonstrated an \textit{in-situ} wavefront-correction approach for two-photon microendoscopy. Since the wavefront correction is sensitive to the bending of the fiber, for each fiber orientation a different correction needs to be found. However, this can be continuously performed during \textit{in-vivo} experiments. In our experimental implementation, the limiting factor on the time required for determining the wavefront-correction was the refresh rate of the specific liquid-crystal SLM used (<5 Hz). This yielded optimization times of tens of minutes in our experiments. This can be significantly shortened by using faster SLMs, since fundamentally the optimization time is limited by the 2PF signal level. In our experiments, integration times of  $0.1[ms]$  were sufficient at the first iterations of the optimization process. Significantly lower integration times are required in the following iterations, when the signal grows. Thus, finding the wavefront correction using SLMs with higher refresh rates, such as Digital Mirror Devices (DMDs)\cite{Conkey2012High-speed}, or galvanometric mirrors based approaches \cite{Papadopoulos2017F-sharp} can decrease the optimization time by more than three orders of magnitudes. Using more advanced algorithms such as genetic algorithms\cite{Conkey2012Genetic} can also reduce the number of iterations required for optimization. These are expected to yield an optimization time of the order of seconds or less even using fluorescent markers, as was recently demonstrated using galvanometric-mirrors for 2PF microscopy through scattering tissue\cite{Papadopoulos2017F-sharp}.  In general, any similar  approaches for wavefront correction using nonlinear feedback that were originally developed for scattering media are directly applicable for bundle-based endoscopy, since a fiber bundle can be considered as a thin scattering layer \cite{porat2016widefield}. After the initial focusing, image acquisition speed is limited by the galvanometric scanners speed. Faster scanners based on acousto-optic scanners, or resonant galvanometric mirrors can be utilized for faster scanning.

An inherent limitation of wavefront-shaping based correction is the sensitivity to fiber bending. While our proximal-only detection approach does not require initial measurements prior to the insertion of the endoscope, or any knowledge of the fiber parameters or shape, any movement of the fiber after the optimization process will hamper the correction and decrease the focus intensity. For a static fiber left untouched, we have experimentally measured decorrelation times of more than 12 hours. However, small bending and shifts of a few millimeters caused complete decorrelation of the speckle patterns and focusing intensity in both fibers used. To overcome this restriction, continuous adaptive focusing may be used to adaptively compensate for the fiber movement during imaging, as small movements of the bundle translates to small shifts and small decorrelation of the focus\cite{warren2016adaptiveEndosMicroscope}. Another possible approach is to deploy the endoscope inside a rigid cannula\cite{Ohayon2018MinimallyInvasive}, which will significantly improve stability at the price of an increase in the overall endoscope diameter, and loss of mechanical flexibility.

The field of view using fiber bundles with ordered cores is limited by the diffraction side-lobes of the periodic arrangement of cores \cite{porat2016widefield}, and the fiber 'memory-effect' angular range. The maximal angular FoV will be attained by using a disordered bundle having  single-mode fibers, with no light propagation between the cores.  Unlike the memory-effect in scattering media, the memory-effect range in fiber bundles is not limited by the fiber length, and for an ideal bundle having single-mode cores, with no coupling, the memory-effect FOV spans the entire NA of the fiber \cite{porat2016widefield}

\section{Conclusion}
\label{Conclusion}
We have demonstrated a  minimally-invasive, lensless, two-photon micro-endoscope with \textit{in-situ} wavefront correction. In contrast to prior works, our technique does not require any distal access or prior characterization, making it an interesting potential solution for imaging or optogenetic stimulation\cite{Valentina2014BundlePhotoactivation}. We used the non-linearity of the two-photon excitation process to generate a focal spot. Other nonlinear mechanisms such as three-photon fluorescence, second harmonic generation, or stimulated Raman scattering may also be used. Using an SLM with faster refresh-rate may allow to use our approach for freely behaving animal studies.

\section*{Acknowledgments}
We thank Valentina Emiliani for fruitful discussions.

\section*{Funding}
We thank the Human Frontiers Science Program (Grant RGP0015/2016), The Israeli Ministry of Science and Technology (Grant 712845), and The Azrieli Foundation. 

\section*{Disclosures}
The authors declare that there are no conflicts of interest related to this article.


\begin{thebibliography}

\bibitem{Vasilis2010GoingDeeper} V. Ntziachristos , "Going deeper than microscopy: the optical imaging frontier in biology", Nat. methods \textbf{7}, 603-614 (2010).
\bibitem{flusberg2005fiberReviewNature} B. A. Flusberg, E. D. Cocker, W. Piyawattanametha, J. C. Jung, E. L. Cheung, and M. J. Schnitzer, '' Fiber-optic fluorescence imaging'',  Nat. methods \textbf{2}, 941 (2005).
\bibitem{oh2013fiberReview} G. Oh, E. Chung, and S. H. Yun, '' Optical fibers for highresolution in vivo microendoscopic fluorescence imaging, '' Opt. Fiber Technol. \textbf{19}, 760-771 (2013).
\bibitem{Sylwia2015Scanning-free} S. M. Kolenderska, O. Katz,  M. Fink, and S. Gigan, "Scanning-free imaging through a single fiber by random spatio-spectral encoding," Opt. Lett. \textbf{40}, 534-537 (2015)
\bibitem{Barankov2014HighThrough-put} R. Barankov, and J. Mertz, "High-throughput imaging of self-luminous objects through a single optical fibre" Nat. communications \textbf{5}, 5581 (2014)
\bibitem{andresen2013twoPhotoneLenslessEndoscope} E. R. Andresen, G. Bouwmans, S. Monneret, and H. Rigneault, '' Two-photon lensless endoscope, '' Opt. express \textbf{21}, 20713-20721 (2013).
\bibitem{Thompson2011Adaptive} A. J. Thompson, C. Paterson, M. A. Neil, C. Dunsby, and Paul M. French, "Adaptive phase compensation for ultracompact laser scanning endomicroscopy," Opt. Lett. \textbf{36}, 1707-1709 (2011)
\bibitem{warren2016adaptiveEndosMicroscope} S. C. Warren, Y. Kim, J. M. Stone, C. Mitchell, J. C. Knight, M. A. Neil, C. Paterson, P. M. French, and C. Dunsby, '' Adaptive multiphoton endomicroscopy through a dynamically deformed multicore optical fiber using proximal detection, '' Opt. express \textbf{24}, 21474-21484 (2016).
\bibitem{vcivzmar2012exploiting} T. {\v{C}}i{\v{z}}m{\'a}r and K. Dholakia, '' Exploiting multimode waveguides for pure fibre-based imaging, '' Nat. communications \textbf{3}, 1027 (2012).
\bibitem{bianchi2012multi} S. Bianchi and R. Di Leonardo, '' A multi-mode fiber probe for holographic micromanipulation and microscopy, '' Lab on a Chip \textbf{12}, 635-639 (2012).
\bibitem{Nenad2008FluorescenceEndomicroscopy} N. Bozinovic,C.  Ventalon, T. Ford and J. Mertz, "Fluorescence endomicroscopy with structured illumination, " Opt. express \textbf{16}, 8016-8025(2008).
\bibitem{papadopoulos2012focusing} I. N. Papadopoulos, S. Farahi, C. Moser, and D. Psaltis, '' Focusing and scanning light through a multimode optical fiber using digital phase conjugation, '' Opt. express \textbf{20}, 10583-10590 (2012).
\bibitem{mosk2012controlling} A. P. Mosk, A. Lagendijk, G. Lerosey, and M. Fink, '' Controlling waves in space and time for imaging and focusing in complex media, '' Nat. photonics \textbf{6}, 283 (2012).
\bibitem{choi2012scanner} Y. Choi, C. Yoon, M. Kim, T. D. Yang, C. Fang-Yen, R. R. Dasari, K. J. Lee, and W. Choi, '' Scanner-free and wide-field endoscopic imaging by using a single multimode optical fiber, '' Phys. review letters \textbf{109}, 203901 (2012)
\bibitem{ploschner2015seeingThroughChaos} M. Pl{\"o}schner, T. Tyc, and T. {\v{C}}i{\v{z}}m{\'a}r, '' Seeing through chaos in multimode fibres, '' Nat. Photonics \textbf{9}, 529 (2015).
\bibitem{Yariv1976Direct} A. Gover, C. P. Lee, and A. Yariv, "Direct transmission of pictorial information in multimode optical fibers, " J. Opt. Soc. Am. \textbf{66}, 306-311 (1976)
\bibitem{Spitz1967Tranmission} E. Spitz, and A. Werts, "Transmission des images {\'a} travers une fibre optique, " Comptes Rendus Hebdomadaires Des Seances De L Academie Des Sciences Serie B, 264 (14), 1015 (1967).
\bibitem{porat2016widefield} A. Porat, E. R. Andresen, H. Rigneault, D. Oron, S. Gigan, and O. Katz, '' Widefield lensless imaging through a fiber bundle via speckle correlations, '' Opt. express \textbf{24}, 16835-16855 (2016).
\bibitem{Nicolino2016CAlibration-free} S. Nicolino, M. Christophe and P. Demetri, "Calibration-free imaging through a multicore fiber using speckle scanning microscopy," Opt. Lett. \textbf{41}, 3078-3081 (2016)
\bibitem{katz2014noninvasive} O. Katz, E. Small, Y. Guan, and Y. Silberberg, '' Noninvasive nonlinear focusing and imaging through strongly scattering turbid layers, '' Optica \textbf{1}, 170-174 (2014).
\bibitem{rosen2015focusing2PFThroughMMF} S. Rosen, D. Gilboa, O. Katz, and Y. Silberberg, '' Focusing and scanning through flexible multimode fibers without access to the distal end, '' arXiv preprint arXiv:1506.08586 (2015).
\bibitem{Vellekop2008PhaseControlAlgorithms} I.M Vellekoop and A.P Mosk, "Phase control algorithms for focusing light through turbid media, " Opt. communications, \textbf{11}, 3071-3080, (2008)
\bibitem{Katz2011FocusingandCompression} O. Katz, E. Small, Y. Bromberg and Y. Silberberg "Focusing and compression of ultrashort pulses through scattering media, " Nat. photonics, \textbf{6}, 372, (2011).
\bibitem{vellekoop2007focusing} I. M. Vellekoop and A. P. Mosk, "Focusing coherent light through opaque strongly scattering media," Opt. Lett. \textbf{32}, 2309-2311 (2007)
\bibitem{sivankutty2016AperiodicFibre} S. Sivankutty, V. Tsvirkun, G. Bouwmans, D. Kogan, D. Oron, E. R. Andresen, and H. Rigneault '' Extended field-of-view in a lensless endoscope using an aperiodic multicore fiber,'' Opt. letters \textbf{41}, 3531-3534 (2016).
\bibitem{kim2018semi} Y. Kim, S. Warren, F. Favero, J. Stone, J. Clegg, M. Neil, C. Paterson, J. Knight, P. French, and C. Dunsby,  ''Semirandom multicore fibre design for adaptive multiphoton endoscopy, '' Opt. express \textbf{26}, 3661-3673 (2018)
\bibitem{Curry2011DirectDetermination} N. Curry, P. Bondareff, M. Leclercq, N.F. van Hulst, R. Sapienza, S. Gigan, and S. Gr{\'e}sillon, "Direct determination of diffusion properties of random media from speckle contrast," Opt. Lett. \textbf{36}, 3332-3334 (2011)
\bibitem{Conkey2012High-speed} D. B. Conkey, A. M. Caravaca-Aguirre, and R. Piestun, "High-speed scattering medium characterization with application to focusing light through turbid media," Opt. express \textbf{20}, 1733-1740 (2012).
\bibitem{Conkey2012Genetic} D. B. Conkey, A. N. Brown, A. M. Caravaca-Aguirre, and R. Piestun, "Genetic algorithm optimization for focusing through turbid media in noisy environments," Opt. express \textbf{20}, 4840-4849 (2012).
\bibitem{Papadopoulos2017F-sharp} I. N. Papadopoulos, J. Jouhanneau, J. A. Poulet, and B. Judkewitz, "Scattering compensation by focus scanning holographic aberration probing (F-SHARP)," Nat. Photonics \textbf{11}, 116-123 (2017).
\bibitem{Ohayon2018MinimallyInvasive} S. Ohayon, A. Caravaca-Aguirre, R. Piestun, and J. J. DiCarlo, "Minimally invasive multimode optical fiber microendoscope for deep brain fluorescence imaging," Biomed. Opt. Express \textbf{9}, 1492-1509 (2018)
\bibitem{Valentina2014BundlePhotoactivation}V. Szabo, C. Ventalon, V. De Sars, J. Bradley, and V.  Emiliani, " Spatially selective holographic photoactivation and functional fluorescence imaging in freely behaving mice with a fiberscope," Neuron, \textbf{84}, 1157-1169.
\end{thebibliography}
\end{document}